\documentstyle[prl,aps,tighten,multicol,epsf]{revtex}

\begin{document}
\begin{multicols}{2}
\narrowtext
{\bf Comment  on ``Anomalous Spreading  of Power-Law
Quantum Wave Packets"}  \\[3mm]

Anomalous spreading of free wave packets with power-law tails
was reported in a recent Letter by Lillo and Mantegna \cite{lm00}.
The authors consider positive even wave functions $\psi(x)$
decreasing with  $x$ as $\psi(x)\sim |x|^{-\alpha}$ ($\alpha>1/2$),
and analyze the asymptotic time evolution through the stationary phase
approximation. They conclude that,  for $\alpha>1$
the well known upper bound $t^{-1/2}$ for the decay of Gaussian and
finite extension packets still stands,  while for $1/2<\alpha<1$ the
decay follows the $t^{-\alpha/2}$ law.
However, they do not give a correct answer for the case $\alpha=1$.
As an illustrative example they work out the special family of wave
functions
\begin{equation} \label{example}
\psi(x)=N\,(x^2+\gamma^2)^{-\alpha/2},
\end{equation}
where $N$ is a normalization constant and $\gamma$ a real parameter.
This family naturally emerges, in fact, in anomalous
diffusion\cite{tsallis}.

We point out that their main result, i.e., the anomalous decay of wave
functions with $|x|^{-\alpha}$ tails was derived before
in Ref. \cite{u98} where a  wave function similar to (\ref{example})
was used as prototype.

A result for general functions decaying as $|x|^{-1}$ is not
provided in \cite{lm00}. However, the authors state that, in the
special case described by Eq. (\ref{example}) and $\alpha=1$,
the maximum of the packet decreases as $1/\sqrt{t}$,
{\it an assertion which is not correct}.

We want to show that there is an extremely direct way to
obtain the {\em exact} decay of packets given by Eq. (\ref{example}).
In particular, it is shown that $\alpha=1$ leads to a $\ln(t)/\sqrt{t}$
decay, as already pointed out in \cite{u97}.
As it is well known, the time evolution of a free wave packet in one
dimension can be described by \cite{qm}
\begin{equation} \label{twave}
\psi(x,t) = \frac{1}{\sqrt{2\pi i\,\hbar t/M}}
\int^{\infty}_{-\infty}
{\rm e}^{\textstyle \frac{ iM \left(x-x^\prime \right)^2}{2\hbar t} }
\psi(x^\prime,0)\; {\rm d} x^\prime,
\end{equation}
where $M$ is the mass of the particle.
Following Ref. \cite{lm00}, we concentrate our attention in the
behavior of the maximum of the wave packet (at $x=0$). Then, we
realize that the integral in Eq. (\ref{twave}) when $\psi(x,0)$
is given by Eq. (\ref{example}) can be evaluated directly by using
the following integral representation of the Kummer function,
$U(\nu, \mu,z)$ \cite{abra}, with $\nu=\frac{1}{2}$ and
$\mu=\frac{3-\alpha}{2}$,
\begin{equation} \label{ww}
U({\textstyle \frac{1}{2}},{\textstyle \frac{3-\alpha}{2}},z)=
\frac{1}{\sqrt{\pi}} \int^{\infty}_{0} {\rm e}^{ -z s}
s^{-1/2} (1+s)^{-\alpha/2} {\rm d}s.
\end{equation}
Hence, $\psi(0,t)$ can be obtained straightforwardly from Eq. (\ref{ww})
after the change of variables $s=(x'/\gamma)^2$ together with the
identification $z=\gamma^2 M/[2i\hbar t]$. In this way we get
\begin{equation} \label{psi0t}
\psi(0,t) = \frac{N\gamma^{1-\alpha}}{\sqrt{2i\hbar t/M}}\;
U({\textstyle \frac{1}{2}},{\textstyle \frac{3-\alpha}{2}},
\gamma^2M/[2i\hbar t] ).
\end{equation}

In order to obtain the asymptotic time behavior of $\psi(0,t)$,
we employ the following limiting forms of the Kummer
function for small $|z|$ \cite{abrb}
\begin{equation} \label{limu}
U({\textstyle \frac{1}{2}},{\textstyle \frac{3-\alpha}{2}},z)
\simeq \left\{
\begin{array}{ll}
\frac{1}{\textstyle \sqrt{\pi}} \Gamma({\textstyle \frac{1-\alpha}{2}})
\;z^{(\alpha-1)/2}   & ({\scriptstyle1/2<\alpha<1}),\\[2mm]
-\frac{1}{\textstyle \sqrt{\pi}}\;\ln(z) & ({\scriptstyle \alpha=1}),
\\[4mm]
\Gamma({\textstyle \frac{\alpha-1}{2}})/
\Gamma({\textstyle \frac{\alpha}{2}})
& ({ \scriptstyle1<\alpha }).
\end{array}
\right.
\end{equation}
Finally, replacing these asymptotic expressions in Eq. (\ref{psi0t})
we get for large $t$
\begin{equation} \label{lim}
|\psi(0,t)| \sim \left\{
\begin{array}{ll}
t^{-\alpha/2}\hspace*{2cm}  & ({\scriptstyle1/2<\alpha<1}),\\[2mm]
\ln(t)\,t^{-1/2}             & ({\scriptstyle\alpha=1}),\\[2mm]
t^{-1/2}                     & ({\scriptstyle1<\alpha}). \\[2mm]
\end{array}
\right.
\end{equation}
So, the $\ln(t)/\sqrt{t}$ law is expected to be the general one governing
the asymptotic evolution of packets decaying as $|x|^{-1}$ since
tails and not other details of the packet shape seem  to rule the
asymptotic behavior.

We acknowledge partial financial support from Brazilian research
agencies FAPERJ and CNPq. \\[4mm]

R. S. Mendes$^{1,2}$ and C. Anteneodo$^3$ \\[3mm] {\small
$^1$Departamento de F\'{\i}sica, Universidade Estadual de
Maring\'a, Av. Colombo 5790, CEP 87290-020, Maring\'a-PR, Brazil,
{\rm e-mail: rsmendes@dfi.uem.br}  \\[3mm] $^2$Department of
Physics, University of North Texas, Denton, TX 76203, USA \\[2mm]
$^3$Instituto de Biof\'{\i}sica, Universidade Federal do Rio de
Janeiro, C. Universit\'aria, CCS, CEP 91949-900, Rio de Janeiro,
Brazil, {\rm e-mail: celia@cat.cbpf.br} } \\[3mm] PACS numbers:
03.65.-w, 05.60.+w

\end{multicols}
\end{document}